\documentclass[aps,prb,twocolumn,superscriptaddress,showpacs]{revtex4-1}
\usepackage{amsmath,amssymb,amsfonts,amsthm}
\usepackage{graphicx}
\usepackage{bm}
\usepackage{bbm}
\usepackage{color}
\usepackage{dcolumn}   
\usepackage{epstopdf}
\usepackage[caption=false]{subfig}
\usepackage{xr}
\usepackage{color}
\usepackage[colorlinks=true,linkcolor=blue,urlcolor=blue,citecolor=blue]{hyperref}

\begin{document}

 \newcommand{\breite}{1.0} 

\newcommand{\beq}{\begin{equation}}
\newcommand{\eeq}{\end{equation}}

\newcommand{\bea}{\begin{eqnarray}}
\newcommand{\eea}{\end{eqnarray}}
\newcommand{\lt}{<}
\newcommand{\gt}{>} 

\newcommand{\Reals}{\mathbb{R}}     
\newcommand{\Com}{\mathbb{C}}       
\newcommand{\Nat}{\mathbb{N}}       

\newcommand{\id}{\mathbboldsymbol{1}}    

\newcommand{\Real}{\mathop{\mathrm{Re}}}
\newcommand{\Imag}{\mathop{\mathrm{Im}}}

\def\O{\mbox{$\mathcal{O}$}}   
\def\B{\mbox{$\mathcal{B}$}}   
\def\C{\mbox{$\mathcal{C}$}}   
\def\F{\mathcal{F}}			
\def\sgn{\text{sgn}}

\newcommand{\ckd}{\ensuremath{c_{k\sigma}^\dag}}
\newcommand{\ck}{\ensuremath{c_{k\sigma}}}
\newcommand{\ep}{\ensuremath{\epsilon_p}}
\newcommand{\deo}{\ensuremath{\Delta_0}}
\newcommand{\Po}{\ensuremath{\ket{\Psi_o}}}
\newcommand{\Pe}{\ensuremath{\ket{\Psi_e}}}
\newcommand{\dea}{\ensuremath{\Delta}}
\newcommand{\aj}{\ensuremath{a_j}}
\newcommand{\ajd}{\ensuremath{a^{\dagger}_{j}}}
\newcommand{\sx}{\ensuremath{\sigma_x}}
\newcommand{\sz}{\ensuremath{\sigma_z}}
\newcommand{\spl}{\ensuremath{\sigma_{+}}}
\newcommand{\smi}{\ensuremath{\sigma_{-}}}
\newcommand{\alk}{\ensuremath{\alpha_{k}}}
\newcommand{\bk}{\ensuremath{\beta_{k}}}
\newcommand{\om}{\ensuremath{\omega}}
\newcommand{\dw}{\ensuremath{\Delta_0}}
\newcommand{\wbp}{\ensuremath{\omega_0}}
\newcommand{\dv}{\ensuremath{\Delta_0}}
\newcommand{\vbp}{\ensuremath{\nu_0}}
\newcommand{\vplus}{\ensuremath{\nu_{+}}}
\newcommand{\vminus}{\ensuremath{\nu_{-}}}
\newcommand{\wplus}{\ensuremath{\omega_{+}}}
\newcommand{\wminus}{\ensuremath{\omega_{-}}}
\newcommand{\uv}[1]{\ensuremath{\mathbf{\hat{#1}}}} 
\newcommand{\dg}{\ensuremath{\dagger}}

\newcommand{\lr}[1]{\left( #1 \right)}
\newcommand{\lrs}[1]{\left( #1 \right)^2}
\newcommand{\lrb}[1]{\left< #1\right>}

\newcommand{\ket}[1]{\left| #1 \right>} 
\newcommand{\bra}[1]{\left< #1 \right|} 
\newcommand{\braket}[2]{\left< #1 \vphantom{#2} \right|
 \left. #2 \vphantom{#1} \right>} 
\newcommand{\matrixel}[3]{\left< #1 \vphantom{#2#3} \right|
 #2 \left| #3 \vphantom{#1#2} \right>} 

\newcommand{\pdd}[2]{\frac{\partial^2 #1}{\partial #2^2}} 
\newcommand{\pdc}[3]{\left( \frac{\partial #1}{\partial #2}
 \right)_{#3}} 
 \renewcommand{\d}[2]{\frac{d #1}{d #2}} 
\newcommand{\dd}[2]{\frac{d^2 #1}{d #2^2}} 
\newcommand{\pd}[2]{\frac{\partial #1}{\partial #2}} 
\newcommand{\grad}[1]{{\nabla} {#1}} 
\let\divsymb=\div 
\renewcommand{\div}[1]{{\nabla} \cdot \boldsymbol{#1}} 
\newcommand{\curl}[1]{{\nabla} \times \boldsymbol{#1}} 
\newcommand{\laplace}[1]{\nabla^2 \boldsymbol{#1}}
\newcommand{\vs}[1]{\boldsymbol{#1}}
\newcommand{\abs}[1]{\left| #1 \right|} 
\newcommand{\avg}[1]{\left< #1 \right>} 

\let\baraccent=\= 

\title{Moir\'e superconductivity}

\author{Ivar Martin}
\affiliation{Material Science Division, Argonne National Laboratory, Argonne, IL 08540, USA}
\email{ivar@anl.gov}

\date{\today}
\begin{abstract}
Recently, superconductivity was discovered at very low densities in slightly misaligned graphene multilayers. Surprisingly, despite extremely low electronic density (about $10^{-4}$ electrons per unit cell), these systems realize strong-coupling superconductivity, with the transition temperature being a large fraction of the Fermi energy ($T_c\sim 0.1 \epsilon_F$). 
Here we propose a qualitative explanation for  this remarkable phenomenon, highlighting similarities and qualitative differences with the conventional uniform high-density superconductivity. 
Most importantly, we find that periodic superimposed potential generically enhances local interactions  relative to nonlocal (for instance, Coulomb) interactions. In addition, the density of states is enhanced as well, exponentially in modulation strength for low lying bands in some cases.
Combination of these two effects makes  moir\'e systems natural  intermediate or strong-coupled superconductors, with potential for very high transition temperatures.

\end{abstract}
\maketitle

\section{Introduction} 

Search for materials with high superconducting transition temperature ($T_c$) has been one of the main quests in physics ever since its discovery  in mercury by Kammerlingh Onnes in 1911. 
The basic enigma of superconductivity was resolved in Bardeen, Cooper, and Schrieffer in 1958\cite{BCS1957} by showing that electron-phonon interaction leads to pairing of itinerant electrons, and that the pair condensation  is responsible for superconductivity. 

While it would seem that stronger interaction  should lead to higher $T_c$, this is not necessarily the case, since  interactions also can make electrons less coherent, which suppresses superconductivity. The proper treatment of intermediate and strong coupling superconductivity, as well as the inclusion of Coulomb repulsion have only become possible after the work of Eliashberg, who, building upon the previous works of Gor'kov\cite{gor1959microscopic} and Migdal\cite{migdal1958interaction}, constructed now famous Dyson's equations for superconductivity \cite{eliashberg1960interactions}. These equations have become de facto {\em the theory}  superconductivity, used  both to interpret experimental observations and to predict new superconductors.

For a long time, the search for high temperature superconductors (HTSC) has been focused on bulk materials obtained by chemical synthesis. Currently, the record holder among bulk materials is LaH$_{10}$\cite{drozdov2019superconductivity, ashcroft2004hydrogen}, which under  megabar pressure becomes a superconductor at 250K.

More recently it has become possible to create a few monolayer 2D materials by MBE \cite{Logvenov699, ge2015superconductivity} or mechanical exfoliation\cite{cao2018unconventional}.  These  methods have opened new ways to control superconductivity that were not available in bulk materials,  including gate doping, tuning strain, dielectric properties, and modification of electronic and phononic states. In a way, there systems realize the old  dream of designer superconductors of Little\cite{Little1964} and Ginzburg\cite{ginzburg1970problem}. 

Perhaps the most unusual method to tune superconductivity to date, demonstrated in 2018, involves creating large (compared to the atomic scale) periodic superstructures in 2D materials. It was first discovered upon stacking two graphene layers with a slight  misalignment  angle, $\theta \approx 1^\circ$("twisted bilayer graphene", or TBG) \cite{cao2018unconventional}.  The misalignment creates a {\em moir\'e} pattern, that has a spatial period that is a factor $1/\theta$ larger than  the atomic unit cell.  Such superstructure leads to Brillouin zone folding into a Brillouin minizone, into which every microscopic electronic band is folded $1/\theta^2$ times.
Remarkably, the observed superconducting $T_c$ can be a few degrees K high, a significant fraction of the Fermi energy within a miniband; the transition temperature is highly sensitive to the angle $\theta$. Since the original discovery, superconductivity has  been observed in other misaligned materials as well \cite{liu2019spin, jauregui2019superconductivity}.  
It should  be noted that moir\'e patterns can  form also when different materials are stacked, with or without angular misalignment \cite{decker2011local}.  This makes it a versatile new way to control superconductivity is layered systems, whose full potential has been barely tapped.

In this paper we  qualitatively analyze electron-phonon interactions, density of states, and Coulomb interactions in moir\'e superstructures and how their interplay affects superconductivity.  Focusing on generic features, our analysis is not limited to graphene moir\'e structures. 

There are several special features that qualitatively distinguish {\em moir\'e superconductors} from their more conventional counterparts. First, the superconducting order parameter acquires an internal moir\'e-scale spatial structure, tracking the spatial modulation  of the electronic wave functions. Second, due to the small width of the minibands, $ w_M$, the frequencies of the phonons (or other bosonic modes) $ \omega_0$ that mediate electron-electron attraction can exceed minibandwidths. As we will show, for $\omega_0> w_M$, the Migdal's criterion, which usually allows to neglect vertex corrections in the Eliashberg equations, is no longer valid. Instead, the ``smallness'' of the vertex corrections becomes controlled by the dimensionless strength of the electron phonon coupling, $\lambda$ (and not  $\lambda \omega_0/\epsilon_F$).
Therefore, only in the case of weak-to-intermediate coupling, $\lambda < 1$,  omitting the vertex corrections can be justified. Finally, when $\omega_0> \epsilon_F$, there is no logarithmic reduction of the Coulomb repulsion that usually occurs in high-density superconductors\cite{bogoliubov1962new, Morel1962}.  Fortunately, we find instead  that long-range Coulomb interaction is suppressed relative to phonon-mediated attraction due to the spatial moir\'e modulation of the electronic wavefunctions.

The rest of the paper is organized as follows. In Section \ref{sec:basics} we summarize the key results of classical theory of superconductivity. In Section \ref{sec:Holst} we show how these results  arise in the Holstein model, highlighting the assumptions made in the standard derivations for high-density superconductors.  We discuss what changes in the case of low density superconductors $\omega_0> \epsilon_F$.
 In Section \ref{sec:moire} we study spatially modulated moir\'e superconductivity. We  show  that starting from the weak-coupling limit, $\lambda \ll 1$, superconductivity in low density systems is generically enhanced compared to the unmodulated case. The reasons is that both electronic density of states (DOS) and phonon-mediated electron-electron attraction are enhanced.  We also show how moir\'e modulation affects differently short range and long range interactions, relatively suppressing long range interactions such as Coulomb.  In Section \ref{sec:carb} we explicitly consider the case of moire twisted graphene systems. Finally, in Section \ref{sec:disc} we discuss results and possible connections between superconductivity in moir\'e and  some other  systems.

\section{The basic principles and results of conventional superconductivity}\label{sec:basics}

In this section we summarize the principal results of conventional superconductivity. Even though the  results were obtained with phonon-based superconductors in mind, they can apply to other pairing mechanism as well, as long as the assumptions (discussed below) are satisfied.

The theory of Bardeen, Cooper and Schrieffer\cite{BCS1957} identified the the key elements that control superconductivity: the electronic density of states near the Fermi level, $\nu_{\epsilon_F}$, the electron-electron attraction induced by phonons, $U$,  and the typical frequency of the relevant phonons, $\omega_0$. Their expression for the transition temperature is
\beq
T_c = \omega_0 e^{-1/\lambda} \label{eq:BCS}.
\eeq
The interaction parameter $\lambda =  \nu_{\epsilon_F} U$ was assumed to be small in  BCS, $\lambda << 1$.
$T_c$ marks the temperature below which the nonlinear self consistent gap  equation 
\beq
1 = U \int_0^{\omega_0} d\epsilon_p \frac{N(\epsilon_p)}{\sqrt{\ep^2 + \Delta^2}}\left[1 - 2n_F(\sqrt{\ep^2 + \Delta^2})\right]\label{eq:Delta}
\eeq
acquires a non-trivial solution, $\Delta \ne 0.$

Physically, the pairing occurs due to electrons polarizing phonons, without actually exciting them out of their ground state. In other words, the pairing interaction is due to  electrons exchanging virtual phonons. This is the origin of the upper cut off  $\omega_0$ in Eq. (\ref{eq:Delta})-- only  electrons with energies within the $\omega_0$ window around the Fermi level $\epsilon_F$ are paired; at higher energies, phonon-induced electron-electron interaction is in fact repulsive \cite{leggett2006quantum}. 
This assumes that $\omega_0 <\epsilon_F$. In the opposite case, the upper cut off in the integral, and thus also the prefactor in Eq. (\ref{eq:BCS}) has to be replaced by $\epsilon_F$.

The simplicity and the elegance of the BCS solution is due to the fact that superconductivity is a weak coupling instability. Thus, it is sufficient to only keep phonon-induced electron-electron interaction, while neglecting the effect of phonons on electron propagation, and vice versa. This is no longer accurate for $\lambda\sim 1$ and above, which is clearly the most interesting regime, since it promises the highest $T_c$ values.

The transition temperature itself is defined only  by the normal state properties of electrons and phonons, and these properties are affected by their mutual interaction\cite{migdal1958interaction, McMillan68}. For electrons, interactions with phonons makes them heavier (polaronic effect), by a factor $1+\lambda$, which leads to an increase in DOS and therefore appears to be good for superconductivity. On the other hand, the quasiparticle residue is reduced by the same factor, and since two electronic Green functions enter the gap equation Eq. (\ref{eq:Delta}), the combined effect of electronic renormalization turns out to be equivalent to $U\to U/(1+\lambda)$, which suppresses $T_c$.
For phonons, the renormalization due to interaction with electrons reduces their frequency by a factor $1-\lambda$. Again, this modification has is a negative effect on superconductivity.

Proper account of these renormalizations as well as of the screened Coulomb interaction is possible within the Eliashberg's   framework\cite{eliashberg1960interactions}. 
Based on numerical solution of the Eliashberg  equations, McMillan  found the following best fit \cite{McMillan68}, for high-density superconductors
\beq
T_c\approx \Theta \exp\left[-\frac{1+\lambda}{\lambda - \mu^*(1+0.6\lambda)}\right] \label{eq:McM}.
\eeq
The characteristic phonon frequency $\Theta$ is an experimentally measurable quantity, and thus includes interaction renormalizations described above.
The apparent reduction of the Coulomb pseudopotential  $\mu^* = \mu/[1 + \mu\ln(\epsilon_F/\omega_0)]$ from the ``bare" (high-frequency value) Coulomb strength $\mu$ occurs due to the smaller frequency range of phonon-induced attraction compared to the Fermi energy \cite{bogoliubov1962new,Morel1962}.
In bulk materials typically $\mu^*$ is between 0 and 0.2\cite{McMillan68}. 

Based on McMillan's formula, the largest $T_c$ would be expected at  values of $\lambda\sim 1$. It should be kept in mind however that increasing $\lambda$ softens phonons, which both reduces $\Theta$ in Eq (\ref{eq:McM}) as well as makes structural instabilities more likely\cite{WhitePhysRevB1992,  Freericks1993}. Even in the absence of structural instabilities,  in a very strong coupling limit one expects electrons to form spatially bound pairs\cite{Alexandrov_2001}, crossing over into BEC regime \cite{chen2005bcs}. 
Empirically, the maximum achievable transition temperature appears to be consistent\cite{WEBB201517, Esterlisnpj} with Eq. (\ref{eq:McM}) taken at $\lambda \approx 1$,  $T_c^{max} \sim 0.1\Theta$.  For instance, in the cases 
of Hg and Pb, where $\lambda \approx 1$, $T_c/\Theta$ is between 0.05 and 0.07 \cite{McMillan68}; similarly high values are reached in Nb compounds and in Ba$_{1-x}$K$_x$BiO$_3$.  Theoretically, for the Holstein model in the $\epsilon_F > \omega_0$ regime, this relationship between maximum achievable $T_c$ and characteristic phonon frequency has been also found by numerical methods distinct from Migdal-Eliashberg approach \cite{Esterlis2018, Hague2008}.

\section{The Holstein model}\label{sec:Holst}

The Holstein model\cite{holstein1959studies} is probably the simplest model that captures the main  features of phonon mediated superconductors. We will use it to illustrate 
the reduction from the full electron-phonon model to an effective BCS Hamiltonian. We will carefully examine the differences between the high-density ($\epsilon_F > \omega_0$) and low-density ($\epsilon_F < \omega_0$) superconductors. The low density result  will be used in the next section were we will consider the case of superconductivity  periodically modulated on large scale. 

The Holstein model assumes local (Einstein) phonons that interact with local electron density. The full Hamiltonian is 
\beq
H = H_{e} + H_{ph} +   \alpha\sum_{j,\sigma} n_{j\sigma} x_j , \label{eq:Hols}
\eeq
with the bare electron and phonon Hamiltonians 
\bea
H_{  e} &=& \sum_k \epsilon_k {\ckd}\ck,\\
H_{ ph} &=& \sum_j \frac{k x_j^2}{2} + \frac{p_j^2}{2M}. \label{eq:H0}
\eea
Here, $k$ is the electron quasimomentum vector, $x_j$ and $p_j$ are  displacement and momentum operators of phonon on site $j$ (frequency $\omega_0^2 = k/M$), electron site occupation number $n_{j\sigma} = c_{j\sigma}^\dag c_{j\sigma}$.

\subsection{Effective Hamiltonian}
An effective purely electronic Hamiltonian can be obtained for electrons whose energies are below the phonon energy $\hbar\omega_0$. For such slow electrons, one can assume that  phonons adjust to the changes in electronic configurations essentially instantaneously. We will demonstrate this reduction procedure both in the first and second quantization. The former has the advantage of greater simplicity, while the latter reveals  the assumptions as well as allows to go beyond the effective Hamiltonian description.

\subsubsection{First quantized (or classical) treatment of phonons}\label{sec:class}
On a single site the part of the Hamiltonian that involves phonon coordinate is $h_i = kx_j^2/2 + \alpha x_j n_j$. Minimizing over the phonon displacement, we find the phonon-induced electron-electron interaction term, $\Delta H_{ee} = - (\alpha^2/k) n_j^2$. 
Since $n_{j\sigma}^2 =n_{j\sigma} $, this term contains a shift of chemical potential, which we will ignore and attractive interaction between electrons, which summed over the full lattice is
\beq
 H_{ee} = - \frac{\alpha^2}{k}\sum_j n_{j\uparrow}n_{j\downarrow}.\label{eq:Hee}
\eeq
Note that the coupling constant 
\beq
U= - \frac{\alpha^2}{k} = -\frac{\alpha^2}{M\omega_0^2}
\eeq
is the static limit of the more general dynamic interaction mediated by phonons\cite{leggett2006quantum}, 
\beq
U(\omega) = \frac{\alpha^2}{M(\omega^2-\omega_0^2)}.
\eeq

\subsubsection{Second quantized treatment of phonons}\label{sec:2ndQ}
We now  show how the same result can be obtained in the second-qauntized language. The phonon Hamiltonian can be expressed in terms of local bosonic operators $b_i = (x_i/\ell_0 + i\ell_0 p_i) $, where the zero-point motion amplitude is given by $\ell_0 = \sqrt{\hbar/M\omega_0}$, as
\bea
H_{ph} &=& \sum_j \omega_0 (b_j^\dag b_j + 1/2),\\
H_{e-ph} &=& \frac{\alpha\ell_0}{\sqrt 2}\sum_{j,\sigma} n_{j\sigma} (b_j^\dag + b_j).\label{eq:Heph}
\eea 
The quantum evolution operator can be expressed in the interaction representation as 
\beq
{\cal U}(t)= e^{-i Ht} = e^{-iH_0t}{\cal T}e^{-i\int_0^t dt' \hat H_{e-ph}(t')},
\eeq
where $\hat H_{e-ph}(t') =  e^{iH_0t}H_{e-ph}e^{-iH_0t}$ and $H_0 = H_e + H_{ph}$.  The effective Hamiltonian can be obtained by the following steps. First, let us expand  the time-ordered exponential up to the second order, 
\beq
 1-i\int_0^t dt' \hat H_{e-ph}(t') - \int_0^t dt' \int_0^{t'} dt'\hat H_{e-ph}(t') \hat H_{e-ph}(t'').\nonumber
\eeq
The next step is averaging over the phonon vacuum $|0\rangle$. This eliminates the linear $b$ in terms. In the second order, only the terms of the form $\bra{0}b_i(t')b^\dag_i(t'')\ket{0} = e^{-i\omega_0(t'-t'')} $ remain.  If electrons are ``slow," (reside within the energy window smaller than $\omega_0$),  then $n_j(t'')$ can be replaced by $n_j(t')$ and integration over $t''$ can be easily performed to give $i/\omega_0$.  The result is 
\begin{widetext}
\beq
1 +\frac{i\ell_0^2 \alpha^2}{2\omega_0}\sum_j\int_0^t dt'[n_j(t')]^2 \approx{\cal T}\exp\left[i \int_0^t dt'\frac{iU}{2}\sum_j[n_j(t')]^2\right]
\label{eq:2ndord}
\eeq
\end{widetext}
Undoing now the interaction representation gives the  Hamiltonian of Eq. (\ref{eq:Hee}). 

This derivation highlights how the band-width limited interaction arises from the Holstein model, and the fact that the BCS Hamiltonian is only accurate to the second order in electron-phonon interaction.
Another assumption  seems to be that phonons must remain in their ground state.  In fact, the same interaction would obtain for any temperature of phonons. This follows from $\int_{-\infty}^0 dt \bra{n}  b^\dagger(0) b(t)+ b(0) b^\dagger(t)\ket{n} = i/\omega_0$, regardless of the phonon number state $\ket{n}$. Strictly speaking, the pairing strength is independent of the phonon temperature, which is a reflection of the fact that phonons are a linear system and thus their differential response does not depend on their state. In equilibrium, this is not important since $T_c$ is lower than $\omega_0$; out of equilibrium, at least in principle, it seems possible to have phonons that are significantly hotter than electrons, and still have pairing. In practice, however, the energy transfer from phonons to electrons would heat electrons up, suppressing superconductivity.

\subsection{Beyond effective Hamiltonian}
For the weak coupling case, $\lambda \ll 1$, the effective BCS  Hamiltonian derived above provides  fully adequate description of  bare electrons experiencing weak mutual attraction within  the frequency window of (bare) $\omega_0$.

Our interest in this work, however, is in the low-density (or narrow band) superconductors, where  $\lambda$ can approach unity. 
While we will not attempt to describe this regime quantitatively, it is worthwhile to pause and asses some  qualitative differences that arise between the  low and high-density superconductors beyond the weak-coupling description of BCS.

\subsubsection{Migdal's criterion }\label{sec:MT}
The Migdal's criterion provides a justification for dropping vertex corrections in the Eliashberg equations.   In conventional high-density superconductors the vertex corrections  are ${\cal O}(\lambda \omega_0/\epsilon_F)$. In the low-density superconductors with   $\omega_0> \epsilon_F$, the original  criterion clearly does not apply and  needs to be  reexamined.

Following Migdal\cite{migdal1958interaction}, the lowest order vertex correction for the Holstein interaction  (\ref{eq:Heph}) is
\beq
\Gamma_1(q) = \alpha^2\ell_0^2 \int_{p'} D(p-p') G(p' + q) G(p') ,
\eeq
where $G = 1/(i\omega - \epsilon_{\vec k})$ and $D = 2\omega_0/(\omega^2 + \omega_0^2)$ are  the bare electronic and phononic Green functions in imaginary time\cite{mahan2013many}. In the vertex equation, for brevity, $p$ denotes both momenta and frequency. $\Gamma_1$ should be compared with the undressed vertex $\Gamma_0 = 1$.
In order to estimate $\Gamma_1$, let's recall that if we omit $D$, the sum over the internal momenta and frequency is the electronic susceptibility 
\beq
\Pi(q) = \int_{p} G(p + q) G(p) = \int_{\vec p}  \frac{n_F(\epsilon_{p+q}) - n_F(\epsilon_{p})}{\epsilon_{p+q} - \epsilon_{p} + i\omega}.\label{eq:Pi}
\eeq
 Both in 2D and 3D, it is typically bounded by the electronic DOS, $\nu_{\epsilon_F}$,  decreasing at large frequencies and momenta.  
Now, for the Holstein phonons, $D$ is only a function frequency. In the case $\epsilon_F > \omega_0$, we can approximately replace $D\sim \epsilon_F^{-1}$. Recalling the definition of the dimensionless coupling constant, that leads to $\Gamma_1\sim \lambda(\omega_0/\epsilon_F)$, the standard Migdal's criterion. In the opposite limit, phonon propagator is simply $D\sim \omega_0^{-1}$, and thus $\Gamma_1\sim \lambda$. 

This is a natural result that says that the vertex corrections can be justifiably neglected for weak coupling superconductivity. In the intermediate coupling regime,  one can only expect  the Migdal-Eliashberg approach  to remain  qualitatively valid.

\subsubsection{Electron propagator renormalization}
Interaction with virtual phonons dresses electronic propagation, transforming electrons into heavier polarons (see Section \ref{sec:basics}). The dressing of low-energy electrons only involves {\em virtual} excitation of phonons. This is an ``off-shell" process that is independent of the ratio between $\epsilon_F$ and $\omega_0$. Thus we expect that the mass renormalization and the quasiparticle residue have the same form both in low and high-density superconductors, with the renormalization factor $1+\lambda.$
 Indeed, the electron self energy, for temperatures below  $\hbar\omega_0$, is
 \bea
\Sigma_{el}(q) &=& \alpha^2\ell_0^2 \int_{p} D(p) G(p + q)\\
&=& \alpha^2\ell_0^2 \int_{\vec p} \frac{n_F(\epsilon_{\vec p})}{i\omega - \epsilon_{\vec p} + \omega_0} + \frac{1- n_F(\epsilon_{\vec p})}{i\omega - \epsilon_{\vec p} - \omega_0}.
\eea
After analytical continuation, the frequency-dependent real part of self energy is
\beq
\Sigma(\omega) =  \alpha^2\ell_0^2 \nu_{\epsilon_F} \log \frac{\omega_0-\omega}{\omega_0+\omega},
\eeq
regardless of the relationship between $\epsilon_F$ and $\omega_0$.  For small frequencies, $|\omega|\ll \omega_0$, $\Sigma(\omega)\approx \lambda \omega$, which gives both the polaronic propagation slowdown and reduction of the  quasiparticle weight. 

\subsubsection{Phonon frequency renormalization}
As mentioned  in Section \ref{sec:basics}, the phonon frequency in  high-density  metals is  reduced due to electron phonon-interactions, $\omega_0\to \omega_0(1-\lambda)$.  This effect can be obtained by considering the  phonon self-energy\cite{migdal1958interaction},
\beq
\Sigma_{ph} (q)= \alpha^2\ell_0^2\int_p G(p+q) G(p) = \alpha^2\ell_0^2 \Pi(q),
\eeq
with the polarization bubble  given by Eq. (\ref{eq:Pi}). 
To obtain phonon frequency renormalization, the electron bubble $\Pi(q)$ has to be evaluated near the phonon frequency.  For high-density superconductors,  $\epsilon_F > \omega_0$,  $\Pi(q)\approx \nu_{\epsilon_F}$. From that immediately follows that $\Sigma_{ph} (q) \approx \lambda\omega_0$, causing the familiar frequency renormalization\cite{migdal1958interaction}, $\omega_0 \to \omega_0(1-\lambda)$. 

However, in the opposite low-density limit, $\epsilon_F < \omega_0$,  $\Pi(\omega_0, \vec q)$ will be strongly suppressed, by a factor which we can estimate as $\epsilon_F/\omega_0$.  Therefore, the frequency renormalization in this case is much weaker, $\omega_0\to\omega_0(1 -  \frac{\epsilon_F}{\omega_0}\lambda )$.
 This is quite natural, since the low-concentration electrons should not be able to strongly renormalize phonon frequencies. 
Note however that this is a welcome change compared to the dense superconductors, where large $\lambda$ needed for high $T_c$, also causes suppression of  phonon frequencies, which opposes the growth of $T_c$.

\section{Enhancement of superconductivity in moir\'e structures}\label{sec:moire}
We now turn to the main subject of this paper -- the effect of periodic supercell modulation on superconducting $T_c$. 
The fact that greatly enhanced  $T_c$ has been experimentally observed in a variety of moir\'e-twisted systems  \cite{cao2018unconventional, liu2019spin, jauregui2019superconductivity} serves as an indication that there may be a general principle at work. With these experiments as an inspiration, in this section, we will discuss the case of  2D.  For the same reason we will be referring to supercell as moir\'e cell. Before we proceed, we would like to note that the effects that lead to $T_c$ enhancement that we identify are not limited to 2D; experimentally creating a 3D moire superstructure, however,  appears to be more challenging.

We will use again the Holstein model. The simple form of electron phonon interactions in the Holstein model
allows to see particularly easily what happens under coarse-graining to a larger unit cell. Despite the simplicity, this model in fact describes accurately general interaction between electrons and longitudinal optical  (LO) phonons, which is probably the most common origin of superconductivity.

We find that periodic moir\'e patterns generically enhance DOS, enhance local attractive electron-electron interaction, and relatively suppress longer large Coulomb repulsion. All of these conspire to enhance $T_c$.

\subsection{Coarse-graining interaction in the Holstein model}

\begin{figure}[htbp]
\includegraphics[width=3.2in]{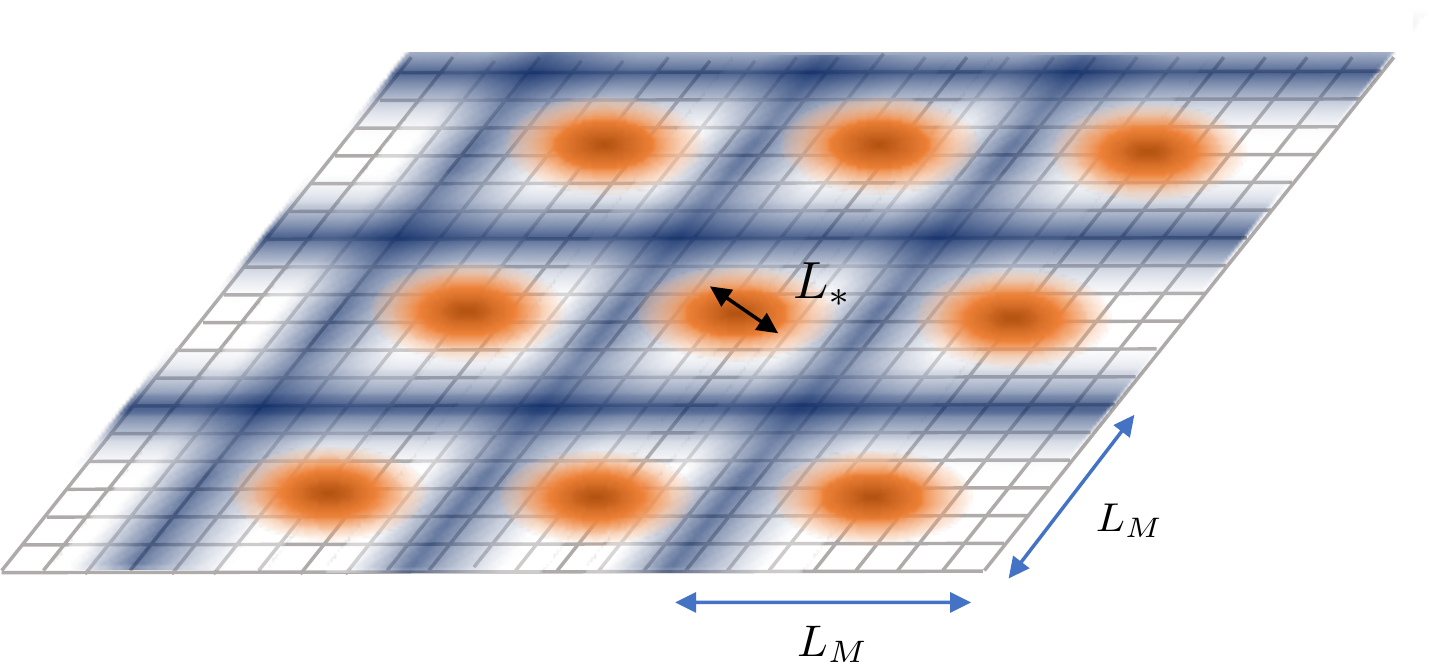}
\caption{A schematic of a  moir\'e lattice. Thin gray lines denote microscopic lattice, with the lattice constant $a_0$. Blue shading represents superimposed moir\'e potential with period $L_M a_0$. Orange spots represent the density of the electronic Bloch functions modulated due to the presence of a strong moir\'e potential. For large supermodulation strength, the width of the density peaks can be significantly smaller that the size of the moir\'e cell, $L_* < L_M$.}
\label{fig:moi}
\end{figure}

Superposing  a periodic structure on an atomic crystal redefines the unit cell (see Figure \ref{fig:moi}). Each original crystal energy band becomes folded $N_M = S_M/S_0$ times, where $S_M = L_M^2a_0^2$ is the real space area of the moir\'e unit cell, and $S_0 = a_0^2$ is  the atomic unit cell area. What does that do to superconductivity? We will address this question using the Holstein model (\ref{eq:Hols}) as a framework, by constructing an effective model,  coarse-grained to the moir\'e cell level.

A low-energy electron within moir\'e cell $J$ (we use capital indices to distinguish from microscopic site indices such as $j$) of the Holstein model resides on $ N_*$ atomic sites and thus interacts with $ N_*$ phonons.  For a weak supercell modulation potential  $N_*\approx N_M$; however, for a strong modulation $N_*$ can be significantly smaller than  $N_M$.  Only $1/N_*$ fraction of the electron interacts with each of $N_*$ phonons. Thus, summed within the moir\'e real space unit cell, the electron-electron interaction becomes
\beq
\frac{\alpha^2}{2k}\sum_{j \in J} n_j^2\approx  \frac{\alpha^2}{2kN_*}n_J^2. 
\eeq
It may appear that coarse-graining to the moir\'e scale strongly reduces interaction relative to Eq. (\ref{eq:Hee}), in proportion to  the number of sites occupied by electrons within the moir\'e cell,
\beq
H_{ee}^M = -\frac{\alpha^2}{2kN_*}\sum_{ J} n_J^2.
\eeq
However, the coarse-graining is accompanied by an increase of coarse-grained DOS. The dimensionless coupling constant $\lambda$, which controls superconductivity, is the product of both of them.

For reference, let's first consider a weak moir\'e potential. Then,  the electronic DOS {\em per moir\'e unit cell} is increased by the factor $N_M$ compared to the DOS per microscopic unit cell,  $\nu_{\epsilon_F}^M \sim \nu_{\epsilon_F} N_M$. Thus, the superconducting  temperature for weak modulation, according to the BCS expression  Eq. (\ref{eq:BCS}) remains approximately the same. This is of course  consistent with the fact that redefining the unit cell without applying any modulation cannot affect superconductivity.

Instead of referring to the coarse-grained moir\'e lattice, it is convenient to refer back to the original microscopic unit cell.  Moving moir\'e factor $N_M$ from DOS to interaction, we see that the interaction is actually  enhanced, 
\beq 
U_{eff} = \frac{N_M}{N_*}U.\label{eq:Ueff}
\eeq
Moreover, the DOS can be also significantly enhanced 
in the strong modulation case. Thus both interactions and DOS conspire to increase $\lambda$ and $T_c$.  

\subsection{Strong moir\'e modulation in a parabolic material}\label{sec:strongmoire}

As an example, let us consider a  2D material, with  lattice constant $a_0$ and  isolated conduction  band of width $W$.  Such a band structure corresponds to the DOS per unit cell of $\nu_{\epsilon_F}\sim 1/W$. 
The effective mass (the curvature near the band bottom) is  $m^*\sim \hbar^2/(a_0^2 W)$ .

Suppose now that a  strong periodic potential  $V(r)$, varying smoothly between $V_{min} $ and $V_{max}$  on scale $L_Ma_0$, is superimposed on the lattice (we pick $V_{min} = 0$ for convenience).
If 
\beq
V_{max}  >   \hbar^2/(a_0^2L_M^2 m^* ) \sim W L_M^{-2}, 
\eeq
 then the electronic wave functions near the bottom of the original band  becomes strongly spatially modulated, peaked at the minima of $V(r)$. This situation is  accounted for by first solving for the intra-well bound states, and then including their inter-well tunneling, which will lead to the formation of low energy narrow minibands.
The band width of the lowest miniband can be estimated as\cite{landau2013quantum}, 
\beq
w_M\sim \sqrt{V_{max} W}\frac{1}{L_M}\exp\left[{-\sqrt{\frac{V_{max}}{W}} {L_M}}\right]
\eeq

The corresponding microscopic DOS (per atomic unit cell) is

\beq
\nu^M_0\sim \frac1{w_M N_M} \sim\frac 1 W \frac{e^\zeta}{\zeta}
\eeq
where 
\beq
\zeta = \sqrt{\frac{V_{max}}{W}} {L_M}>1. 
\eeq

In addition, in this regime, due to the wave function concentration near the minima of $V(r)$, $N_*$ is significantly smaller than $N_M$; for the lowest bands it can be estimated as $N_*\sim N_M/\zeta$, and therefore, from Eq. (\ref{eq:Ueff})
\beq
U_{eff}\approx \zeta U.
\eeq

Finally, we obtain that in the strong modulation limit, the dimensionless phonon-mediated electron-electron attraction is enhanced relative to the unmodulated value exponentially,
\beq
\lambda_M^{deep} \approx e^\zeta \lambda_0.
\eeq
Neglecting for the moment the electronic Coulomb repulsion, this is the coupling constant that determines the superconducting transition temperature for deeply modulated moir\'e structure. For $\lambda_M < 1$,
\beq
T_c^M\sim \min(\epsilon_F,\omega_0) e^{-1/\lambda_M^{deep}}.\label{eq:TcM}
\eeq
The prefactor is modified compared to the standard BCS to allow for  large phonon frequencies, $\omega_0>\epsilon_F$.
While in deriving this expression we used the moir\'e renormalization of interaction derived within the Holstein model, as we will show in the next section, the result is more general.

\subsection{Short range vs. long rage interactions in moir\'e systems}
Above we found that deeply modulated superlattices can have strongly enhanced tendency to superconduct due to the 
increased values of DOS and phonon-mediated attraction. We haven't addressed however the problem of Coulomb interaction, which opposes phonon attraction, and also can be expected to become modified under superlattice modulation. In fact, it is obvious that an onsite Coulomb repulsion (Hubbard type), would transform identically to the purely local phonon-mediated attraction in the Holstein model. Thus, for local interactions, the balance between attraction and repulsion will remain unchanged under moir\'e modulation.

The physical Coulomb repulsion is however not local. How will the nonlocality affect the way Coulomb transforms in the presence of  the moir\'e modulation?
In this section, by more carefully deriving the renormalization of interaction potential we find that longer range interactions are {\em enhanced less} than the short range interactions. We therefore generally expect that  supercell modulation {\em suppresses} Coulomb interaction relative to the phonon-mediated attraction.

Let us consider a generic interaction Hamiltonian,
\beq
H_{int} = \sum_{r, r'}v(r - r') n_r n_{r'}.
\eeq
For $v(r) \propto \delta_{r,r'}$ it corresponds to the Holstein model with phonons integrated out  (electron energies within $\omega_0$ from each other).  The onsite densities, $n_r = \sum_\sigma n_{r\sigma}  = \sum_\sigma \psi_\sigma^\dag(r)\psi_\sigma(r)$, can be expressed in terms of the Bloch mode functions, $\psi^\dag_\sigma(r) = \sum_{q, m}e^{iqr}u_{q,m}(r)c^\dag_{q,m}$, where $u_{q,m}(r)$ is the moir\'e-periodic part of the single electron wave function, quasimomentum $q$ lies within the moir\'e (folded) Brillouin zone, $m$ is moir\'e miniband index, and operator $c^\dag_{q,m}$ creates electron in that mode.

Suppose we are only interested in the interactions within one miniband. Dropping the minband index,  the Hamiltonian becomes
\begin{widetext}
\begin{equation}
H_{int} = \sum_{r, r'}v(r - r') e^{i(k - k')r+i(p - p')r'}c^\dag_{k\sigma}c_{k'\sigma}c^\dag_{p\sigma'}c_{p'\sigma'} f_{kk'}(r) f_{pp'}(r'),
\end{equation}
\end{widetext}
where $ f_{kk'}(r) = u_k(r)u^*_{k'}(r)$ is moir\'e-periodic. For deep minibands $m$, $u_k(r)$ will be nearly independent of $k$, and thus $f^m_{kk'}(r) \approx |u_{k=0,m}(r)|^2 \equiv \rho^m(r)$ is the single electron density (scales as $1/N$, where $N$ is the  number of sites in the whole system).  It is a periodic function, and thus can be expanded in harmonics 
$$\rho(r) = \frac{1}N\sum_G e^{i Gr}\rho_G,$$
where $G = mG_1 + nG_2$ are the lattice points of 2D  moir\'e reciprocal lattice.  By construction, $\rho_{G = 0} = 1$.  In the absence of moir\'e modulation this is the only non-zero $\rho_G$.  However, if the density $\rho(r)$ is modulated, as it is for deep moir\'e lattice, the number of $G$s for which $\rho_{G } \sim 1$ is given by the ratio of the real space moir\'e cell  to the area of the support of $u(r)$ within the cell. That is $N_M/N_* = L_M^2/L_*^2$.

Substituting this into the interaction Hamiltonian, we find,

\begin{widetext}
\begin{equation}
H_{int} = \frac 1{N^2}\sum_{r, r'}v(r - r') e^{i(k - k' + G)r+i(p - p' + G')r'}c^\dag_{k\sigma}c_{k'\sigma}c^\dag_{p\sigma'}c_{p'\sigma'} \rho_{G} \rho_{G'},
\end{equation}
\end{widetext}

Requiring now that the scattering leaves both electrons in the same moir\'e Brillouin minizone implies $G + G' = 0$, and keeping only the terms in the Cooper channel we find
\begin{equation}
H_{Cooper} = \frac 1{N}\sum_{k,p,G, \sigma, \sigma'}\tilde v_{k - p+G} c^\dag_{k\sigma}c^\dag_{-k\sigma'}c_{-p\sigma'} c_{p\sigma}\ \rho_{G} ^2.
\end{equation}

For short range interaction, $\tilde v_q$ is momentum independent and thus we see the moir\'e enhancement of interaction by the factor $\sum_G\rho_{G} ^2\approx N_M/N_*$.  Note that this result does not require that $v(r)$ is atomic-scale local: same enhancement applies as long as the range of $v(r)$ is less than $L^* = \sqrt{N_*}$. 
If interaction has a longer range, $L_{int}>L_*$, then the amplification factor is reduced to $L_M^2/L_{int}^2.$ Truly long range interaction, such as unscreened Coulomb,  is not enhanced by the moir\'e modulation at all. This will be critical when we discuss twisted bilayer graphene in the next section.

\section{Carbon superconductivity}\label{sec:carb}
In this section we present  estimates of the key parameters that determine the viability of the phonon origin of superconductivity in twisted bilayer  \cite{cao2018unconventional, lu2019superconductors} and  double bilayer\cite{liu2019spin} graphene. 
The analysis is again qualitative, focusing on absolute and relative strengths of phonon-mediated attraction and Coulomb repulsion, ignoring the precise details of materials and band structure. The conclusion that we reach is that the phonon-mediated pairing interaction strength, combined with the high density of states in these systems are sufficient to make them $\lambda\sim 1$, intermediate coupling superconductors. As discussed in Section \ref{sec:strongmoire}, under such conditions $T_c$ can reach about $10\%$ of Debye frequency or Fermi energy, whichever is less.
(In twisted graphene structures, the miniband widths are a few meV, much smaller than the characteristic phonon frequency of about 100 meV).

Following the logic of Section \ref{sec:Holst}, we start from the ``parent" untwisted, but highly doped, graphene and then turn to the twisted moir\'e system. 
This allows to contrast clearly how the low-density moire system is related but different from its highly doped uniform parent.

\subsection{Doped graphene}

A lot is known about the form of electron phonon interaction in graphene\cite{BA2008}.  Qualitatively, however, the coupling between electrons and LO phonons can be easily estimated by noting that the primary effect of carbon-carbon bond length on itinerant  electrons is to change the overlap between   $\pi$ orbitals.  The local electron-phonon interaction can then be approximated by the Holstein coupling of Eq. (\ref{eq:Hols}), with $\alpha\approx t_\pi/a_0$ and the mass of the Einstein phonon by the carbon mass, $M\approx M_C$.  
This leads to the following estimate of the phonon-mediated attraction strength between electrons 
\beq
U\sim \frac{(t_\pi/a_0)^2}{M_C \omega_0^2}.
\eeq
Here  $\omega_0$ is a typical LO phonon frequency, which is of the order of 100 meV. 
An estimate for this interaction, given that $t_\pi/\hbar\omega_0\sim 30$, is about an eV. In combination with the band width of several eV, this gives  $\lambda$ in the intermediate coupling range.

To determine feasibility of superconductivity, we need  to compare this attraction strength with the strength of Coulomb repulsion.
In the Fourier space, the phonon-mediated interaction is flat,
\beq
V_q^{ph} \sim -\frac{t_\pi^2}{M_C \omega_0^2}.
\eeq

The Coulomb interaction in stand-alone graphene at charge neutrality point is unscreened,  $V(r) = e^2/\epsilon r$, where $\epsilon$ is the dielectric constant of embedding medium. In doped graphene, the screening length becomes finite, $q_{sc}\sim k_F$. \cite{Castro2009}  In the Fourier space this leads to 
\beq
V_q^{C} \sim \frac{e^2}{\epsilon(q + q_{sc})}.
\eeq
In the presence of a metallic gate distance $d$ away from graphene, $q_{sc} \sim {\rm max} (1/d, k_F)$. 

Qualitatively, we expect that if phonon mediated attraction  dominates, then there is a good chance for phonon-mediated superconductivity. The ratio of the two interactions is
\beq
\left|\frac{V_q^{ph}}{V_q^C}\right|\sim\epsilon \left(\frac{t_\pi}{\hbar \omega_0}\right)^2\frac{m_e}{M_C} q_{sc} a_0.\label{eq:balance}
\eeq
In this estimate we ignored  the distinction between the carbon mass and the reduced mass of the oscillator, lattice constant $a_0$ and the Bohr radius and other order 1 constants, focusing on the parametric dependencies.
In the heavy doping regime, $q_{sc} a_0\sim1$. Taking  $t_\pi/\hbar\omega_0\sim 30$ and $\epsilon\sim 10$, we find that the two interactions are indeed comparable, even without pseudopotential renormalization\cite{Morel1962}. 

\subsection{Twisted moir\'e graphene}
In the twisted graphene devices, the density of carriers is very low, $L_M^{-2}\approx 10^{-4} $ electrons per atomic unit cell. In some devices, there is also a gate distance $d$ away, which can be comparable to $L_Ma_0$. This implies that $q_{sc}\sim 1/(L_Ma_0)$, which would make the ratio of phonon to Coulomb interactions in Eq. (\ref{eq:balance}) tiny!

However, with the moir\'e amplification, as we saw in Section \ref{sec:strongmoire}, the short range interactions are enhanced by the factor $(L_M/L_*)^2$, while Coulomb, screened on distances $L_M$, is not! Thus, the moir\'e version of Eq. (\ref{eq:balance})  is

\beq
\left|\frac{V_q^{ph}}{V_q^C}\right|_{moire}\sim\epsilon \left(\frac{t_\pi}{\hbar \omega_0}\right)^2\frac{m_e}{M_C} \frac{a_0 L_M}{L_*^2}. 
\eeq
This new interaction ratio can become again order 1, e.g. if $L_M\sim 100 a_0$ and $L_* \sim 10 a_0$.  This is sufficiently close to the theoretically expected value of $L_*$,\cite{bistritzer2011moire, WMM2018, PhysRevResearch.1.033072} supporting the point that  phonon induced interaction remains competitive and may even overcome Coulomb repulsion in moir\'e structures  despite very weak screening. 

Attractive overall  interaction  by itself does not guarantee superconductivity at a  reasonable temperature. However, at the magic angle, not only the attraction is enhanced, but also the electronic DOS at low energies is much higher than in pristine graphene. 
The nearly flat electronic minibands that appear near the magic angle can have badwidths $w_M$ of a few meV. This corresponds to DOS$\sim 1/(w_M L_M^2)$ which happens to be of the same order as the DOS in heavily doped graphene, $1/(t_\pi a_0^2)$. 

As we saw in the previous section,  for heavily doped graphene $\lambda$ is in the intermediate coupling regime. With the enhanced interactions in moir\'e twisted graphene, $\lambda$ can easily reach intermediate or even strong coupling. In this case, as discussed in Section \ref{sec:strongmoire}, $T_c$ can be as high as  $0.1 \min(\omega_0, \epsilon_F)\to 0.1 \epsilon_F$, which is consistent with the experimental observations of a few Kelvin $T_c$'s. 

\subsection{Caveats}
In this section we focused on the energetics in an attempt to see whether electron-phonon coupling in graphene  has enough bare strength and spatial structure to dominate Coulomb repulsion and yield a reasonable $T_c$. It appears that it does.
This however does not rule out other types of correlated physics, particularly given the observed proximity to many commensurate insulating phases, usually attributed to Mott physics\cite{lu2019superconductors}.

Even within the phonon scenario, there are many peculiar features that are invisible to the coarse approach that we took. More careful consideration reveals that superconductivity mediated by phonons is likely to be in the d-wave channel\cite{WMM2018}, and can be topologically nontrivial \cite{PhysRevB.99.195114}. It is also interesting to remark that the superconducting order parameter in moir\'e systems is highly inhomogeneous, reminiscent of Josephson junction arrays. One should keep in mind however, that the similarity is superficial, since each grain on average contains at most one Cooper pair, and thus it is impossible to define a phase associated with the ``island." That does not prevent, however,  the possibly having an interesting order parameter structure {\em intra-moir\'e unit cell}.

\section{Discussion}\label{sec:disc}

In this paper we qualitatively studied how superconductivity is affected by a large-scale periodic modulation superimosed on top of the periodic atomic potential. This work was inspired by the recent observation of superconductivity at extremely low electronic densities in  twisted multilayers of graphene\cite{cao2018unconventional, liu2019spin, lu2019superconductors} and transition metal dichalcogenides\cite{jauregui2019superconductivity}. 

Small relative twists  between layers lead to long period moir\'e potentials that can have non-perturbative effect on electronic wavefunctions.
Instead of trying to capture detailed physics of  particular systems, we attempted to more broadly examine qualitative effects of such large scale structures on superconductivity. We have found that quite generally, the effect of supermodulations on superconductivity is positive: both local attractive interactions and  electronic density of states are enhanced, both leading to an increase in the dimensionless electron phonon coupling constant $\lambda$. Moreover, the phonon softening that usually accompanies strong coupling limit and negatively affects $T_c$ is reduced in the low density limit.

So it would appear that moir\'e modulation is an overall excellent way to enhance superconductivity. There is a trade-off however:   low electronic density in moir\'e systems  implies low superfluid density. This not only entails lower critical current, but also can limit $T_c.$
Indeed, in 2D systems, superfluid density, and hence the density itself,  controls the Berezinsky-Kosterlitz-Thouless (BKT) transition temperature\cite{berezinsky1970destruction,kosterlitz1973ordering}.  
General analysis of optical sum rules\cite{PhysRevX.9.031049} leads to the  bound on $T_c$ for electrons in parabolic band, $T_c<\epsilon_F/8$: Small $\epsilon_F$ necessarily limits the allowed  $T_c$, even if $T_c/\epsilon_F$ can be large. 
The same bound also follows from Eq. (\ref{eq:TcM}). 
In order to  optimize the {\em absolute value} of $T_c$, systems need to be tuned to the regime where  $\lambda \sim 1$, while $\epsilon_F$ is still large.

It is interesting to note a parallel between moir\'e superconductors and cuprates.
Based on  experimental\cite{PhysRevLett.62.2317} evidence and theoretical reasons \cite{emery1995importance}, it is likely that the peak of superconducting transition temperature as a function of doping in cuprates occurs at the crossover between the BCS and BKT (phase fluctuations dominated)  regimes.  The connection between moir\'e systems and cuprates may be even closer, than  superficial comparison of the phase diagrams of two systems  suggest. In many cuprates,  charge {\em stripes}\cite{berg2009striped} may be providing an effective superstructure, similar to moir\'e. This supermodulation is also capable of locally increasing paring interaction at the expense of reduced superfluid stiffness \cite{PhysRevB.72.060502}. 

There is also a tantalizing connection between moir\'e superconductivity and the negative-$U$ center mechanism for pairing proposed by Anderson\cite{PhysRevLett.34.953}. This  is one of the leading candidate mechanisms for superconductivity\cite{geballe2016paired} in  lightly Tl doped semiconductor PbTe\cite{Ka_danov_1985}.  For strong moir\'e supermodulation, due to the  suppression of Coulomb repulsion relative to the phonon-mediated attraction, a single moire-cell can play a role similar to the negative-$U$ center,  pairing electrons within itself. In contrast to the doped semiconductors, where pairing is conjectured to occur on the randomly distributed  valence-skipping dopants, the moir\'e cells are ordered in real space, and the minibands are much narrower than the phonon frequency. This alleviates some of the main concerns\cite{geballe2016paired} that were expressed with regards to the applicability of the negative-$U$ mechanism to the  doped PbTe (and related) systems. It is possible that moir\'e twisted superconductors may represent a clean realization of the Anderson's idea.

Using moir\'e twist is one of the most radical new approaches to tune material properties. Even though there are limitations (e.g., the trade offs between $T_c$ and superfluid density), it is clear that the ability to accurately impose supercell structure provides a powerful additional  knob that can be added to the existing repertoire of chemical, structural, mechanical and other means for controlling materials. It is quite likely that this additional flexibility  may allow  to construct synthetic material with properties that are hard or impossible to reach otherwise, including room temperature superconductivity.

\acknowledgments
Author would like to acknowledge many fruitful discussions with Fengcheng Wu, Mohammad Hafezi, Mike Norman.  
This work  was supported by the Materials Sciences and
Engineering Division, Basic Energy Sciences, Office of
Science, U.S. Dept. of Energy.

\bibliographystyle{apsrev4-1}
\bibliography{Eliash}

\end{document}